\documentclass[12pt]{article}
\usepackage{amssymb}
\title{On the admissible families of  components  of \\ Hamming codes}
\author{Alexander M. Romanov \\
\small Sobolev Institute of Mathematics \\[-0.8ex]
\small 630090 Novosibirsk, Russia\\
\small {rom@math.nsc.ru}\\
}
\usepackage{hyperref}
\date{}
\begin{document}
\maketitle

\begin{abstract}
In this paper, we describe the properties of  the $i$-components of Hamming codes. We suggest  constructions of the admissible families of components of Hamming codes. It is shown that every $q$-ary code of length $m$ and minimum distance 5 (for $q = 3$ the minimum distance is 3)  can be embedded  in a  $q$-ary  1-perfect code of length $n = (q^{m}-1)/(q-1)$. It is also shown that every binary code of length $m + k$ and minimum distance
$3k + 3$  can be embedded  in a  binary 1-perfect code of length $n = 2^{m}-1$.
\end{abstract}


\section{Introduction \label{Properties}}

Let $\mathbb{F}_{q}^{n}$ be a vector space of dimension $n$ over the Galois field ${\mathbb{F}}_{q}$. The Hamming distance between two vectors ${\bf x}$, ${\bf y} \in \mathbb{F}_{q}^{n}$ is the number of coordinates in which they differ and it is denote by $d({\bf x}, {\bf y})$. An arbitrary subset  ${C}$ of  $\mathbb{F}_{q}^{n}$ is called $q$-ary  {\it 1-perfect} code of length $n$, if for every vector ${\bf x} \in \mathbb{F} _{q}^{n}$ there exists a unique vector $ {\bf c} \in {C}$ such that $d({\bf x}, {\bf c}) \leq 1$. It is known that $q$-ary 1-perfect  codes of length $n$   exist only if $n = (q ^{m} -1) / (q-1)$, where $m$ is a natural number not less than two. We shall assume that the all-zero vector $ {\vec 0} $ is in code. A code is called {\it linear} if it is a linear space over ${\mathbb{F}}_{q}$. The  linear 1-perfect codes  are called {\it Hamming} codes. The $q$-ary Hamming code of length $n$ is denoted by $\mathbb{H}$.

The {\it weight} of a vector ${\vec x} \in \mathbb{F}_{q}^{n}$ is the number $d({\vec x},{\vec 0})$. A vector of weight $3$ of the  code $\mathbb{H}$ is called {\it triple}. Consider the subspace $R_i$ spanned by the set of all triples of the code $\mathbb{H}$  having  $1$ in the  $i$-th coordinate. All  cosets $R_i + {\vec u}$  form the set of  {\it $i$-components} of the $q$-ary  Hamming code $ \mathbb {H}$,  where $i \in \{ 1,2 \dots n \}$,  ${\vec u} \in  \mathbb{H}$. A family of  components $R_{i_1} + {\vec u}_1,  R_{i_2} + {\vec u}_2, \dots , R_{i_t} + {\vec u}_t$ of the $q$-ary  Hamming code $ \mathbb {H}$ is called {\it admissible} if for  $r, s \in \{ 1, 2, \dots, t \}$, $r \ne s$, we have $(R_{i_{r}} + {\vec u}_{r}) \cap (R_{i_{s}} + {\vec u}_{s}) = {\varnothing}$. See \cite {rom2}.

Let  $n_1 \leq n_2$, $C_1 \subseteq  \mathbb{F}_{q}^{n_1}$ è $C_2 \subseteq  \mathbb{F}_{q}^{n_2}$. We lengthen all the vectors of the code $C_1$ to the length of $n_2$ by appending a zero vector of length $n_2 - n_1$. They say that the code $C_1$ can be {\it embedded} in the code $ C_2 $ if all the lengthened  vectors of $C_1$ belong to $C_2$. We consider all the vectors of the code $C_2$  in which the last $n_2 - n_1$ coordinates are equal to zero. We delete the last $ n_2 - n_1 $ coordinate in all such vectors. If the resulting set of shortened vectors coincides with $C_1$, then we say that the code $C_1$ can be embedded in the code $C_2$ in the strong sense.

Avgustinovich and Krotov \cite {avg}  showed that any binary  code of length $m$ and minimum distance $3$ can be embedded (in the strong sense) in a binary 1-perfect code of length $2^m -1$.

In this paper, we describe  properties of  the $i$-components of Hamming codes. We suggest  constructions of the admissible families of components of Hamming codes. It is shown that every $q$-ary code of length $m$ and minimum distance 5 (for $q = 3 $ the minimum distance is 3)  can be embedded  in a  $q$-ary  1-perfect code of length
$n = (q^{m}-1)/(q-1)$. It is also shown that every binary code of length $m + k$ and distance $3k + 3$  can be embedded  in a  binary 1-perfect code of length $n = 2^{m}-1$.

We present three examples of the admissible families of components of Hamming codes.  In Example \hyperref[Example1]{1} , for an arbitrary $q$-ary code $(\Lambda \cup \{{\vec 0} \}) \subset \mathbb {F}_{q}^{m}$ with minimum distance 5 we construct an admissible family of components of $q$-ary Hamming code of length $n = (q^m-1)/(q-1)$. The admissible family of component is constructed so that  switching the components of this family, we obtain a $q$-ary 1-perfect code ${\mathbb T}$ of length $n$ in which can be embedded the $q$-ary code $\Lambda \cup \{{\vec 0} \}$ of length $m$. In Example \hyperref[Example2]{2}, for an arbitrary ternary code of length $m$ and distance 3 in exactly the same method as in Example \hyperref[Example1]{1} we  constructing  an admissible family of components of the ternary Hamming code of length $n = (3^m - 1)/2$.
In Example \hyperref[Example3]{3}, for an arbitrary binary code of length $m + k$ and distance $3k + 3$ we  constructing an admissible family of component  of the binary Hamming code of length $n = 2^{m}-1$. The  admissible families of components from Examples \hyperref[Example2]{2} and \hyperref[Example3]{3} have the same properties as the admissible family of components from Example \hyperref[Example1]{1} and allow us, by switching   the components, to  construct the 1-perfect codes in which can be embedded codes of smaller length.

In Section \hyperref[Properties]{2} we present theorems describing the properties of the $i$-components of code ${\mathbb H}$.
In Section \hyperref[Example1]{3} we describe the constructions of admissible families of components of code ${\mathbb H}$.
In Section \hyperref[Example23]{4} we give Examples 2 and 3.
In Section \hyperref[Embedding]{5} we prove a theorem on the embeddability.

The parity-check matrix $H$ of the code $\mathbb {H}$ of length $n = (q^{m} - 1)/(q-1)$ consists of $n$ pairwise linearly independent column  vectors ${\vec h}_i$.
The transposed column vector ${\vec h}_i$ belongs to $\mathbb {F}_{q}^{m}$, $i \in \{1, \dots, n \}$. We assume that the columns of the parity-check matrix $H$ are arranged in some fixed order. The set  $\mathbb{F}_{q}^{m} \setminus \{ \vec 0 \}$ generates a projective space $PG_{m-1} (q)$ of dimension $(m-1)$ over the Galois field  ${\mathbb{F}}_{q}$. In this space, points correspond to the columns of the parity-check matrix  $H$ and the three points $i, j, k$ lie on the same line if the corresponding columns  $\vec h_i, \vec h_j, \vec h_k$ are linearly dependent.
We denote by $l_{xy}$ the line passing through the points $x$ and $y$, and we denote by $P_{xyz}$ the plane spanned by three non-collinear points $x, y, z$. Let ${\vec x} = (x_1, x_2, \dots, x_n) \in \mathbb{F}_{q}^{n}$. Then,  the {\it support} of the vector ${\vec x}$ is the set $ supp {({\vec x})} = \{ i : x_i \neq 0 \}$.
A triple belongs to the line if the support of this triple belongs to the line. The triples intersect at the point $i$ if their supports intersect at the point $i$.

\section{Properties of $i$-components \label{Properties}}

Next, we present theorems describing the properties of the $i$-components of code ${\mathbb H}$.
Let the  subcode $\mathbb{H}_l$ of code $\mathbb{H}$ be defined by the line  $l$. We consider the pencil of lines $l_1, l_2, \dots, l_{(n-1)/q}$ through a point $i$. It is known \cite {rom1} that

\begin{equation}
R_i = \mathbb{H}_{l_1} +  \mathbb{H}_{l_2} + \cdots + \mathbb{H}_{l_{(n-1)/q}}. \label{eq1}
\end{equation}

{\bf T\,h\,e\,o\,r\,e\,m\, 1. }\label{the1}
{ \it Let a vector $ {\vec u} = (u_1, u_2, \dots, u_n) \in R_i$ and a component $u_x$ of the vector $\vec u$ be nonzero, $x \neq i$. Then on the line $l_{ix}$ there exists a point $y$ distinct from the points $i, x$ and such that component $u_y$ of the vector $\vec u$ is nonzero. }

{\bf P\,r\,o\,o\,f.} The basis of the subspace $R_i$ is formed by all linearly independent triples of the code
$\mathbb {H}$ having  $1$ in the  $i$-th coordinate. Consider representation of the vector $\vec u$ with respect to the basis.
From the conditions of the theorem, it follows that in this representation  is a triple whose support contains points $i, x$  and a point which is on the line $l_{ix}$ and is distinct from the points $i, x$. From formula \hyperref[eq1]{1}, it follows that the basis  triples belonging to the line $l_{ix}$ form a subspace  $\mathbb {H}_{l_{ix}}$.
The basis triples, that belong to other lines from the pencil  of lines containing the point $i$, intersect with the basis triples, that lie on the line $l_{ix}$, only at one point $i$. The theorem is proved.

{\bf T\,h\,e\,o\,r\,e\,m\, 2. } \label{the2}
{ \it Let $i \neq j$, a vector ${\vec u} = (u_1, u_2, \dots, u_n) \in R_i + R_j$, a component $u_x$ of the vector $\vec u$ be nonzero and the point $x$ does not lie on $l_{ij}$. Then on the plane $P_{ijx}$ there exists a point  $y$ distinct from the points $i, j, x$ and such that component $u_y$ of the vector $\vec u$ is nonzero.}

{\bf P\,r\,o\,o\,f.}
This theorem is proved similarly to the previous one. The basis triples of $R_i + R_j$ that  lie on the plane $P_{ijx}$ form a subspace. The lines from the pencil of lines  containing the point $i$ either lie on the plane $P_{ijx}$ or intersect with this plane at only one point $i$. The lines of the  pencil of lines through the point $j$ have the same property. The theorem is proved.

\section{Example 1 \label{Example1} }

Next, we describe the constructions of admissible families of components of code ${\mathbb H}$.

{\bf E\,x\,a\,m\,p\,l\,e\, 1. }

In the parity-check matrix $H$ of the Hamming code ${\mathbb H}$ of length $n = (q^{m} -1)/(q-1)$, we choose $m$ linearly independent columns. We assume that we have chosen the columns  $\vec h_1, \vec h_2, \dots, \vec h_m$.
Let  $(\Lambda \cup \{{\vec 0} \}) \subset \mathbb {F}_{q}^{m}$ be a code containing $t$ nonzero vectors ${\vec \lambda}_1, {\vec \lambda}_2, \dots, {\vec \lambda}_t$, the weight of each of them be greater than or equal to three. Let the distance between any two distinct vectors from the set $\Lambda = \{{\vec \lambda}_1, {\vec \lambda}_2, \dots, {\vec \lambda}_t \}$ be greater than or equal to  five. With each vector ${\vec \lambda}_s = ({\lambda}_{s1}, {\lambda}_{s2}, \dots, {\lambda}_{sm})$ of length $m$ we associate a vector ${\vec u}_s$ of length $n$, where $s \in \{1, \dots, t \}$. Let

$$
\mu_s  \vec h_{i_s} = \lambda_{s1}  \vec h_1 +  \lambda_{s2}  \vec h_2 + \cdots +  \lambda_{sm}  \vec h_m,
$$
where $ \mu_s \in \mathbb{F}_q $, $ i_s \in \{1, 2, \dots, n \} $.   Then we put
$$
{\vec u}_s = ({\lambda}_{s1}, {\lambda}_{s2}, \dots, {\lambda}_{sm}, 0, \dots, 0, -\mu_s, 0, \dots, 0).
$$

The support of the vector ${\vec u}_s$ belongs to $\{1, 2, \dots, m \} \cup \{i_s \}$. Since the Hamming code
$\mathbb {H}$ forms a null space of parity check matrix $H$, we have ${\vec u}_s \in \mathbb {H}$.
Thus, based on vectors of length $m$ from the set $\Lambda$, we  constructed a family of components $R_{i_1} + {\vec u}_1, R_{i_2} + {\vec u}_2, \dots, R_{i_t} + {\vec u}_t$ of the $q$-ary Hamming code $\mathbb {H}$ of length $n =(q^{m} -1) / (q-1)$.

Etzion and Vardy \cite {etz} used a set of linearly independent columns of the parity-check matrix of the Hamming code for constructing the full-rank binary 1-perfect codes.

Next, we show that the family of components from Example \hyperref[Example1]{1}  is admissible.

{\bf P\,r\,o\,p\,o\,s\,i\,t\,i\,o\,n  1.} \label{prop1}

{ \it Let $s \in \{ 1, 2, \dots, t\}$. Then,
$ {\vec u_s} \notin   R_{i_s}$.}

{\bf P\,r\,o\,o\,f.}
From the construction it follows that the support of the vector $\vec u_s = (u_1, u_2, \dots, u_n)$ belongs to $\{1, 2, \dots, m \} \cup \{i_s \}$ and column $\vec h_{i_s}$ is a linear combination of three or more columns from the set $\{\vec h_1, \vec h_2, \dots, \vec h_m \}$. Since the columns  $\vec h_1, \vec h_2, \dots, \vec h_m$ are linearly independent, it follows that for $x \in \{1, 2, \dots, m \}$ no linear combination of columns  $\vec h_{i_s}$ and ${\vec h}_x$ does not belong to $ \{\vec h_1, \vec h_2, \dots, \vec h_m \} \setminus \{\vec h_x \}$.
Thus from Theorem \hyperref[the1]{1} we have that ${\vec u_s} \notin R_{i_s}$. The  proposition is proved.

{\bf T\,h\,e\,o\,r\,e\,m\, 3. } \label{the3}

{ \it The family of the component $R_ {i_1} + {\vec u}_1, R_{i_2} + {\vec u}_2, \dots, R_{i_t} + {\vec u}_t$ of the $q$-ary Hamming code $\mathbb {H}$ of length $n$ is admissible.}

{\bf P\,r\,o\,o\,f.}
Let $r, s \in \{ 1, 2, \dots, t \}$, $r \ne s$.
Then, we show that
\begin{equation}
(R_{i_{r}} + {\vec u}_{r}) \cap (R_{i_{s}} + {\vec u}_{s}) = {\varnothing}.   \label{eq2}
\end{equation}
In order to satisfy  the equality \hyperref[eq2]{2}, it suffices to show that $\vec u_r - \vec u_s \notin R_{i_r} +R_{i_s}$.
We consider several cases.

1.
Let $i_r = i_s$.

Then, the vectors $\vec u_r$ and $\vec u_s$ are linearly dependent. From the construction of vectors $\vec u_r$ and $\vec u_s$ implies that the weight of vector $\vec u_r - \vec u_s$ is greater than or equal to six. Hence, arguing as in the proof of Proposition \hyperref[prop1]{1}, we obtain that $ \vec u_r - \vec u_s \notin R_{i_r}$.

2.
Let $i_r \neq i_s$.

Then, we show that $\vec u_r - \vec u_s \notin R_{i_r} + R_{i_s}$. By Theorem \hyperref[the2]{2}, it suffices to show that  the support of  vector $ \vec u_r - \vec u_s $ contains a point $x$ not lying on the line $l_{i_r i_s}$ and such that no other point (distinct from the points  $ i_r, i_s, x $) of the support does not belong to the plane $P_{i_ri_sx}$.

2.1.
Let the columns  $\vec h_{i_r}$ and $\vec h_{i_s}$ be such that as a result of any  linear combination of these columns, one  obtains  a column, which can be represented linear combination of three or more columns from the set $\{\vec h_1, \vec h_2, \dots, \vec h_m \}$.

The support of  vector $\vec u_r - \vec u_s$ belongs to $\{1, 2, \dots, m \} \cup \{i_r \} \cup \{i_s \}$.
Consequently, the point $x$ belong to $\{1, 2, \dots, m \}$. Since the columns  $ \vec h_1, \vec h_2, \dots, \vec h_m $ are linearly independent, it is obvious that none of the columns of the set $\{\vec h_1, \vec h_2, \dots, \vec h_m \} \setminus \{\vec h_x \}$  is not a linear combination of the columns  $  \vec h_{i_r}, \vec h_{i_s},  \vec h_x$. Consequently, $ \vec u_r - \vec u_s \notin R_{i_r} + R_{i_s}$.

2.2.
Let the columns  $\vec h_{i_r}$ and $\vec h_{i_s}$ be such that as a result of  linear combination of these columns, one  obtains  the column $\vec h$ which can be represented linear combination of  two  columns $\vec h_{y'}, \vec h_{y''}$ from the set $\{\vec h_1, \vec h_2, \dots, \vec h_m \}$.

2.2.1.
If at least one of the points $y', y''$ do not belong to the support of  vector $\vec u_r - \vec u_s$,
then $\vec u_r - \vec u_s \notin R_{i_r} + R_{i_s}$.

2.2.2.
Let  $y ', y''$ be the points belonging to the support of  vector $\vec u_r - \vec u_s$.
Then, we choose a point in the support of the vector  $\vec u_r - \vec u_s$ which is distinct from the points $ i_r, i_s, y', y''$. Such a choice is possible by the construction of vectors $ \vec u_r, \vec u_s$. The support of the vector $\vec u_r - \vec u_s$ belongs to $\{1, 2, \dots, m \} \cup \{i_r \} \cup \{i_s \}$. Hence, the points $x, y ', y''$ belong to $\{1, 2, \dots, m \}$ and are not collinear. Consider any other linear combination of columns  $ \vec h_{i_r}, \vec h_{i_s}$  as a result of which, we obtain  a column that is linearly independent from the column $h$. Since the distance between the vectors $ \vec \lambda_r$ and $ \vec \lambda_s$ greater than or equal to five and columns $ \vec h_1, \vec h_2, \dots, \vec h_m $ are linearly independent, it follows that the distance between the columns  $ \vec h_r $ and $ \vec h_s $ is also greater than or equal to five.
Consequently, as a result of the linear combination, we obtain a column which is a linear combination of three or more columns from the set $ \{\vec h_1, \vec h_2, \dots, \vec h_m \}$. Thus, $ \vec u_r - \vec u_s \notin R_{i_r} + R_{i_s}$.

2.3.
Let the columns  $ \vec h_{i_r}$ and $ \vec h_{i_s}$ be such that as a result of a linear combination of these columns, we obtain the column from the set $ \{\vec h_1, \vec h_2, \dots, \vec h_m \}$.
Then the same arguments as in the previous case, we  obtain that $ \vec u_r - \vec u_s \notin R_{i_r} + R_{i_s}$.
The theorem is proved.

\section{Examples 2 and 3  \label{Example23}}

Next, we give Examples 2 and 3. The family of components in these examples are constructed in exactly the same way as in Example \hyperref[Example1]{1}.

{\bf E\,x\,a\,m\,p\,l\,e\, 2. } \label{Example2}

Let $(\Lambda \cup \{{\vec 0} \}) \subset \mathbb {F}_{3}^{m}$ bet a code  containing $t$ nonzero vectors
$ {\vec \lambda}_1, {\vec \lambda}_2 \dots, {\vec \lambda}_t$, the weight of each of them be greater than or equal to three. Let also the distance between any two distinct vectors in $ \Lambda = \{ {\vec \lambda}_1, {\vec \lambda}_2 \dots, {\vec \lambda}_t \}$ be greater than or equal to three. Then, the set $ \Lambda$ corresponds to admissible family of component of the ternary Hamming code of length $ n = (3^m - 1) / 2$.

{\bf E\,x\,a\,m\,p\,l\,e\, 3. } \label{Example3}

In the parity-check matrix $H$ of the binary Hamming code ${\mathbb H}$ of length $n = 2^m -1$, we choose $m$ linearly independent columns. We assume that we have chosen the columns  $\vec h_1, \vec h_2, \dots, \vec h_m$.
In the parity-check matrix $H$, we also choose $k$ columns which  are linear combination of  two columns from  $\{ \vec h_1, \vec h_2, \dots , \vec h_m \}$. We assume that these columns are the columns  $ \vec h_{m + 1}, \vec h_{m + 2}, \dots, \vec h_{m + k}$. Let  $ (\Lambda \cup \{{\vec 0} \}) \subset \mathbb {F}_{2}^{m + k}$ be a code contains $t$ nonzero vectors ${\vec \lambda}_1, {\vec \lambda}_2, \dots, {\vec \lambda}_t$, the weight of each of them be greater than or equal to $3k + 3$.  Let the distance between any two distinct vectors in
$\Lambda =  \{ {\vec \lambda}_1, {\vec \lambda}_2 \dots, {\vec \lambda}_t \}$ be greater than or equal to $ 3k + 3$. With eeach vector ${\vec \lambda}_s = ({\lambda}_{s1}, {\lambda}_{s2}, \dots, {\lambda}_{sm + k})$ of length $m + k$ same way as in Example \hyperref[Example1]{1}, we associate a vector ${\vec u}_s$ of length $n = 2^m -1$, where $s \in \{1, \dots, t \}$. Then, the set $\Lambda $ corresponds to  admissible family of components of the binary Hamming code of length $ n= 2^m -1$.

The proof of the fact that  the families  of components in Examples \hyperref[Example2]{2} and \hyperref[Example3]{3} are admissible is similar to the proof of Theorem \hyperref[the3]{3}. In the case of ternary codes,  we should take in account the features of the Galois field $ \mathbb F_{3}$. Let $\vec x$, $\vec y \in \mathbb {F}_{3}^{m}$. Then, it is obvious that if $d(\vec x, \vec y) = m$ and vectors $\vec x $, $ \vec y$ does not contain zero components, then they are linearly dependent.

\section{Embedding in  perfect code  \label{Embedding}}

Next, we prove a theorem on the embeddability.

By ${\vec e}_i$  we denote a vector of length $n$, where $i$-th component is equal to $1$ and other components are equal to $0$.

Let \label{eq3}
\begin{equation}
 {\mathbb T} = \left (  \mathbb{H} \setminus \bigcup_{s=1}^t  (R_{i_s} + {\vec u}_s)  \right )  \cup \left (\bigcup_{s=1}^t  (R_{i_s} + {\vec  u}_s + \mu_s \cdot {\vec e}_{i_s})  \right).
\end{equation}

By Theorem \hyperref[the3]{3}, the family of component $R_{i_1} + {\vec u}_1, R_{i_2} + {\vec u}_2, \dots, R_{i_t} + {\vec u}_t$ of $q$-ary Hamming code $ \mathbb {H}$ is  admissible (similar  theorems on the admissibility of the family component  are valid for the codes from Examples \hyperref[Example2]{2} and \hyperref[Example3]{3}). Consequently, the set ${\mathbb T}$ is $q$-ary 1-perfect code of length $n$, see \cite {etz, ph}. By Proposition \hyperref[prop1]{1}, the code ${\mathbb T}$ contains the zero vector.

{\bf T\,h\,e\,o\,r\,e\,m\, 4. }

{ \it Every $q$-ary code of length $m$ and minimum distance 5 (for $q = 3$ the minimum distance is 3)  can be embedded  in a  $q$-ary  1-perfect code of length $n = (q^{m}-1)/(q-1)$. Every binary code of length $m + k$ and  minimum distance $3k + 3$  can be embedded  in a  binary 1-perfect code of length $n = 2^{m}-1$.}

{\bf P\,r\,o\,o\,f.}
From the construction of the admissible family of component of $q$-ary code $ {\mathbb H}$ in Example \hyperref[Example1]{1} and formula \hyperref[eq3]{3}, it follows that every $q$-ary code $ \Lambda \cup \{{\vec 0} \}$ of length $m$ and minimum distance 5 can be embedded (in the strong  sense) in the $ q $-ary 1-perfect code ${\mathbb T}$ of length
$n = (q^{m}-1) / (q-1)$.

From the construction of  the admissible family  of component of  ternary code $ {\mathbb H}$ in Example \hyperref[Example2]{2} and Formula \hyperref[eq3]{3}, it follows that every ternary code $ \Lambda \cup \{{\vec 0} \}$ of length $m$ and minimum distance 3 can be embedded (in the strong sense) in a ternary 1-perfect code ${\mathbb T}$ of length $n = (3^m - 1) / 2$.

From the construction of admissible family of component of binary  code ${\mathbb H}$ in Example \hyperref[Example3]{3} and Formula  \hyperref[eq3]{3}, it follows that every binary code $\Lambda \cup \{{\vec 0} \}$ of length $m + k$ and minimum distance $3k + 3$ can be embedded in a binary 1-perfect code ${\mathbb T}$ of length $n = 2^{m}-1$, $k \geq 0$.
In the case of binary codes of Example \hyperref[Example3]{3} the embedding is not strong.
The theorem is proved.

 \end{document}